\documentclass{phyeauth}

\usepackage{graphicx}
\usepackage{amsmath}
\usepackage{amssymb}
\usepackage{subfigure}
\usepackage{bm}
\usepackage{bbm}
\usepackage{stmaryrd} 
\usepackage{psfrag}

\newcommand{\op}[1]{\hat{#1}}
\newcommand*{\bra}[1]{\mathopen{\langle}#1\mathclose{|}}
\newcommand*{\ket}[1]{\mathopen{|}#1\mathclose{\rangle}}
\newcommand*{\Haver}[1]{\mathopen{\llbracket} #1 \mathclose{\rrbracket}}
\newcommand{\Energie}[3][]{E^{#2}_{#3} #1}
\newcommand*{\erwart}[1]{\mathopen{\langle} #1 \mathclose{\rangle}}
\newcommand{\Hgas}{\Hamiltonian^{\text{g}}}
\newcommand{\Hcon}{\Hamiltonian^{\text{c}}}
\newcommand{\ie}{i.e.}
\newcommand{\hilbertspace}{Hilbert space}
\newcommand{\EnergyG}[1][]{\Energie{\text{g}}{#1}}
\newcommand{\EnergyC}[1][]{\Energie{\text{c}}{#1}}
\newcommand{\Kommutator}[2]{\big[#1,#2\big]}
\newcommand{\Hamiltonian}{\hat{H}}
\newcommand{\WW}{\hat{I}}
\newcommand{\Dgas}[4]{\ket{#1,#2}\bra{#3,#4}}
\newcommand{\wkurzBei}{\wahr{}{AB}}
\newcommand*{\Abs}[1]{\left|#1\right|}
\newcommand{\PsiElem}[4]{\psi_{#1 #2}^{#3 #4}}
\newcommand{\productstate}{product state}
\newcommand{\wahr}[3][]{W^{#2}_{#3} #1}
\newcommand{\wkurzGas}{\wahr{}{A}}
\newcommand{\wkurzCon}{\wahr{}{B}}
\newcommand{\qmarks}[1]{``#1"}
\newcommand{\purity}{P}
\newcommand*{\Pmin}{\purity_{\text{min}}^{\text{g}}}
\newcommand{\Pgas}{\purity^{\text{g}}}
\newcommand{\dichtematrix}{\hat{\rho}}
\newcommand{\Dmin}{\dichtematrix^{\,\text{g}}_{\text{min}}}
\newcommand{\refsec}[1]{Sect.~\ref{#1}}
\newcommand{\reffig}[1]{Fig.~\ref{#1}}

\newcommand{\Entartung}[3][]{N^{\text{#2}}_{#3} #1}
\newcommand{\EntkurzGas}{\Entartung{}{A}}
\newcommand{\EntkurzCon}{\Entartung{}{B}}
\newcommand{\hsav}{\hilbertspace\ average}
\newcommand{\tr}[2][]{\text{Tr}_{#1}\left\{#2\right\}}
\newcommand{\precondition}{pre-condition}
\newcommand{\densityoperator}{density operator}
\newcommand{\Entropy}{S}
\newcommand*{\Smax}{\Entropy^{\text{g}}_{\text{max}}}
\newcommand{\Sgas}{\Entropy^{\text{g}}}

\newcommand{\kBolz}{k_{\mathrm{B}}}
\newcommand{\subsystem}{subsystem}
\newcommand{\Energy}{E}
\newcommand{\WahrscheinlichkeitEnergy}{\wahr[(\Energy)]{}{}}
\newcommand{\EnergyGas}[1][A]{\Energie{\text{g}}{#1}}
\newcommand{\EnergyCon}[1][B]{\Energie{\text{c}}{#1}}
\newcommand{\energyconservation}{energy conservation}
\newcommand{\wkurzDom}{\wahr{\text{d}}{AB}}
\newcommand{\cf}{cf.}
\newcommand{\dgas}{\dichtematrix^{\,\text{g}}}
\newcommand{\wkurzDomGas}{\wahr{\text{d}}{A}}
\newcommand{\wkurzDomCon}{\wahr{\text{d}}{B}}
\newcommand{\EntGas}[1][A]{\Entartung[({\EnergyGas[#1]})]{g}{}}
\newcommand{\EntCon}[1][B]{\Entartung[({\EnergyCon[#1]})]{c}{}}
\newcommand{\EntEnergy}{\Entartung[(E)]{}{}}
\newcommand{\NAB}{\EntkurzGas \EntkurzCon}
\newcommand{\Ent}[1]{\Entartung{#1}{}}
\newcommand{\gsystem}{gas system}
\newcommand{\etc}{etc.}
\newcommand{\Deq}{\dichtematrix^{\,\text{g}}_{\text{eq}}}
\newcommand{\konstzwei}{N^{\text{c}}_0}
\newcommand{\konsteins}{\alpha}
\newcommand{\E}{\text{e}}
\newcommand{\Temp}{T}
\newcommand{\pop}[2]{\frac{\partial #1}{\partial #2}}
\newcommand{\ZDichte}[2][]{G^{\text{#1}}(#2)}
\newcommand{\hamiltonian}{Hamiltonian}
\newcommand{\nonequilibrium}{non-equilibrium}
\newcommand{\name}[1]{#1}
\newcommand{\wahrwn}[1]{\wahr{}{#1}}
\newcommand{\Entn}[1]{\Entartung{}{#1}}
\newcommand{\Energyn}[1]{\Energie{}{#1}}
\newcommand{\Funktion}[2]{#1\kern-0.2em\left(#2\right)}
\newcommand{\expfkt}[1]{\Funktion{\exp}{#1}}
\newcommand{\multilevel}{multi-level}
\newcommand{\Emax}{\Energie{}{\text{max}}}
\newcommand{\dd}{\text{d}}
\newcommand{\dod}[2]{\frac{\dd #1}{\dd #2}}
\newcommand{\microstate}{micro state}
\newcommand{\D}{\text{d}}
\newcommand{\Wahrscheinlichkeit}[1]{\wahr[({#1})]{}{}}
\newcommand{\bipartite}{bipartite}
\newcommand{\nondegenerate}{non-degenerate}
\newcommand{\masterequation}{master equation}
\newcommand{\twolevel}{two-level}
\newcommand{\threelevel}{three-level}

\newcommand{\fivelevel}{five-level}
\newcommand{\twolevelsystem}{\twolevel\ system}
\newcommand{\Gaussian}{\name{Gauss}ian}
\newcommand{\sgleichung}{\name{Schr\"odinger} equation}
\newcommand{\threelevelsystem}{\threelevel\ system}

\newcommand{\fivelevelsystem}{\fivelevel\ system}
\newcommand{\WahrscheinlichkeitDomGas}[1][A]
           {\wahr[({\EnergyGas[#1]})]{\text{d}}{}}

\newcommand{\WahrscheinlichkeitGas}[1][A]{\wahr[({\EnergyGas[#1]})]{}{}}

\newcommand{\setup}{set-up}
\newcommand{\densitymatrix}{density matrix}

\begin{document} 

%
%
\begin{frontmatter}

\title{Emergence of thermodynamic behavior within composite quantum systems} 

\author[adr1]{G\"unter Mahler\corauthref{cor}},
\corauth[cor]{Corresponding author. Tel.: +49 711 685 5101;
                                    Fax: +49 711 685 4909.}
\ead{mahler@theo1.physik.uni-stuttgart.de (G.Mahler)}
\author[adr2]{Jochen Gemmer} and
\author[adr1]{Mathias Michel}

\address[adr1]{Institut f\"ur Theoretische Physik I, Universit\"at Stuttgart, 70550 Stuttgart, Germany}
\address[adr2]{Fachbereich Physik, Universit\"at Osnabr\"uck, 49069 Osnabr\"uck, Germany}

\date{\today}%
\begin{abstract}
Entanglement within a given device provides a potential resource for quantum 
information processing. Entanglement between system and environment leads
to decoherence (thus suppressing non-classical features within the system)
but also opens up a route to robust and universal control. The latter is 
related to thermodynamic equilibrium, a generic behavior of bi-partite quantum 
systems. Fingerprints of this equilibrium behavior (including relaxation and
stability) show up already far from
the thermodynamic limit, where a complete solution of the underlying
Schr\"odinger dynamics of the total system is still feasible.
\end{abstract}

\end{frontmatter}

%
%

%
%
\section{Introduction}
\label{chap:qt}

There have been various attempts to reduce thermodynamics to some underlying more fundamental theory. While the vast 
majority of the pertinent work done in this field has been based on classical mechanics \cite{Biggus2002}, a reduction to quantum mechanics
has also attracted increasing interest \cite{Saito1996}.
Decoherence \cite{Giulini1996} has, during the last years, often been discussed as one of the main obstacles for the implementation of
large-scale quantum computation. Quantum thermodynamics \cite{Gemmer2004}, on the other hand,
tries to show that decoherence is far from being just a technical nuisance but a generic phenomenon of partite quantum 
systems giving rise to some of the most dominating, if classical, features of closed (finite) quantum systems: 
thermal equilibrium.

The program of a quantum foundation of thermodynamics should contain the following points:\\
(1) a definition of thermodynamic quantities (in the ideal case, as a function of microstates); (2) a derivation of the
second law of thermodynamics under appropriate constraints (including stability, irreversibility, universality); (3) a
justification of the Gibbsian fundamental form (state functions and conjugate variables); (4) a proof of extensivity or
intensivity, respectively, of the thermodynamic variables; (5) a characterization of thermodynamic systems 
(as opposed to other systems); (6) a kind of correspondence principle (explaining the efficiency of
standard classical approaches despite their underlying quantum nature).

Here we cannot do justice to this rather challenging program. Instead we want to address some of the main
results available to us.

%
%
\section{The Model}

We consider a bipartite system -- an observed system or gas g and an environment or container c, described
by the Hamiltonian
\begin{equation}
  \hat{H} = \hat{H}^{\text{g}} + \hat{H}^{\text{c}} + \hat{I}\;.
\end{equation}
Note that also the environment requires a full quantum treatment and should not be replaced by (classical) boundary
conditions.
$\hat{I}$ defines the interaction between these two subsystems.

Weak coupling between system and environment has routinely to be assumed in standard thermodynamics \cite{Diu1989}; otherwise the
concept of intensive and extensive variables would lose its meaning. Furthermore, in this case the
full spectrum of the coupled system will not look significantly different from the one that results from a mere
convolution of the two spectra of the uncoupled system.
To quantify the weak coupling pre-condition, we require
\begin{equation}
  \sqrt{\erwart{\hat{I}^2}} \ll \erwart{\Hgas} , \; \erwart{\Hcon} \;.
\end{equation}
This inequality must hold for all states that the total system can possibly evolve into under given constraints.

The weak coupling has further to be classified.
Not so much for practical, but for theoretical reasons, the most important contact conditions are the 
\emph{microcanonical} and the \emph{canonical} conditions.
In the microcanonical contact scenario no energy transfer between system and environment is allowed, as opposed to the 
canonical contact.

%
%
\section{Microcanonical Conditions}
\label{chap:qt:sec:microcan}

If a system is thermally isolated, it is not necessarily isolated in the microscopic sense, \ie, not uncoupled to any other system. 
The only constraint is that the interaction with the environment should not give rise to any energy exchange.
As will be seen later, this does not mean that such an interaction has no effect on the considered system, a fact that might seem counterintuitive from a classical point of view.
This constraint, however, leads to an immense reduction of the region in \hilbertspace\ which the wave vector is allowed to enter. 
This reduced area is called \qmarks{accessible region} of the system.

%
%
\subsection{Accessible Region (AR)}
\label{chap:qt:sec:microcan:1}

If the energies contained in the gas g and the environment c, respectively,
\begin{equation}
  \label{eq:40}
  \EnergyG := \erwart{\Hgas}\;,\quad 
  \EnergyC := \erwart{\Hcon}
\end{equation}
are to be conserved, i.e. if these two energies are constants of motion, the following commutator relations should hold
\begin{equation}
  \label{eq:41}
  \Kommutator{\Hgas}{\Hamiltonian}=0\;,\quad %
  \Kommutator{\Hcon}{\Hamiltonian}=0\;.
\end{equation}
It then follows from
\begin{equation}
  \label{eq:97}
  \Kommutator{\Hgas}{\Hamiltonian} = %
  \Kommutator{\Hgas}{\Hgas} + %
  \Kommutator{\Hgas}{\Hcon} + %
  \Kommutator{\Hgas}{\WW} = 0 
\end{equation}
that
\begin{equation}
  \label{eq:42}
  \Kommutator{\Hgas}{\WW}=0\;, \quad %
  \Kommutator{\Hcon}{\WW}=0\;.
\end{equation}
Except for these constraints we need not specify $\WW$ in more detail. 
All interactions that fulfill this relation will create perfectly microcanonical situations, regardless of their strength or any other feature. 
And, as will be shown, there are a lot of possible interactions that do fulfill these conditions and create entanglement and therefore give rise to the increase of local entropy.

Due to (\ref{eq:41}) the local energy projectors $\op{P}^{\text{g}}_A$ of the gas system and $\op{P}^{\text{c}}_B$ of the container
\begin{equation}
  \label{eq:682}
  \op{P}^{\text{g}}_A = \sum_a \Dgas{A}{a}{A}{a}\;, \quad %
  \op{P}^{\text{c}}_B = \sum_b \Dgas{B}{b}{B}{b}
\end{equation} 
commute with the full Hamiltonian,
\begin{equation}
  \label{eq:683}
  \Kommutator{\op{P}^{\text{g}}_A}{\Hamiltonian} = %
  \Kommutator{\op{P}^{\text{c}}_B}{\Hamiltonian} = 0\;.
\end{equation} 
Here $a(b)$ specify the degenerate states (degeneracies $N_A$($N_B$)) for given energy eigenvalue $A(B)$ in the gas (container) 
system.
Thus, because the system is not allowed to exchange energy with the environment the joint 
probability $\wkurzBei$ must be conserved
\begin{align}
  \label{eq:157}
  \bra{\psi}\op{P}^{\text{g}}_A\op{P}^{\text{c}}_B\ket{\psi} %
  &= \sum_{a,b}\Abs{\PsiElem{a}{b}{A}{B}(t)}^2 %
   = \sum_{a,b}\Abs{\PsiElem{a}{b}{A}{B}(0)}^2 \notag\\
  &= \wkurzBei\;,
\end{align} 
and is set by the initial state.
This means that the energy probability distribution $\{\wkurzBei\}$ is a constant of motion.
Vice versa, any state that features this same energy probability distribution as the initial state belongs to the accessible region and could possibly be reached during microcanonical dynamics.

In the following we mainly consider initial {\productstate}s, states that have zero local entropy in the beginning and for which
\begin{equation}
  \label{eq:44}
  \sum_{a,b}\Abs{\PsiElem{a}{b}{A}{B}(0)}^2 = %
  \sum_{a,b}\Abs{\PsiElem{a}{}{A}{}(0)}^2 \Abs{\PsiElem{}{b}{}{B}(0)}^2 %
  = \wkurzGas\wkurzCon \;.
\end{equation}
This is the only constraint that microcanonical conditions impose on the accessible region of \hilbertspace.
Note that this does not mean that local entropy is constant.

%
%
\subsection{The \qmarks{Landscape} of $\Pgas$ in the Accessible Region}
\label{chap:qt:sec:microcan:2}

To demonstrate that the largest part of the accessible region 
is filled with states of almost minimum purity (maximum entropy), we proceed as follows:

\begin{enumerate}

\item First we compute the (unique) state with the lowest possible purity, $\Dmin$ (with purity $P(\Dmin)=\Pmin$) that is 
consistent with the given initial state and the microcanonical conditions, consequently with a given energy probability 
distribution $\{\wkurzGas\}$.

\item Then we compute the average of $\Pgas$ over the total accessible 
\hilbertspace\ region, denoted by $\Haver{\Pgas}$.
 
\item We show that this average purity is very close to the purity of the lowest possible purity 
state $\Dmin$ for a large class of systems.  
Considering only these systems, which then define the class of thermodynamic systems, we can conclude 
that $\Pgas\approx \Pmin$ for almost all states within the accessible region.
Note that this conclusion is only possible because of the fact that the purity of $\Dmin$ is the 
absolute minimal purity which can be reached at all in the system. 
A quantity with a mean value close to a boundary cannot vary very much.
Thus it is not possible that the distribution of the purity within the accessible region is something else but a very flat \qmarks{lowland}, with a \qmarks{soft depression} at $\Dmin$ (see \reffig{fig:purland}) and a \qmarks{peak} with $\Pgas = 1$. 

\begin{figure}
\centering
\psfrag{1}{1}
\psfrag{0}{0}
\psfrag{peq1}{$P = 1$}
\psfrag{empty}{}
\psfrag{p_rest}{$P \approx \Pmin$}
\psfrag{p_min}{$\Pmin$}
\psfrag{rmin}{$\Dmin$}
\psfrag{AR}{Accessible Region}
\includegraphics[width=5cm]{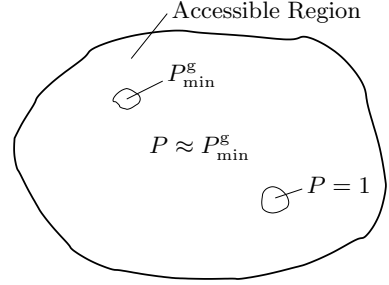}
\caption{Qualitative picture of the purity landscape in the microcanonical case. The biggest part of the accessible region  is at $P \approx \Pmin$ or at $P = \Pmin$. There is however only a small zone featuring $P$ significantly above $\Pmin$ or at the extreme $P = 1$. The only topological property this rough picture refers to is the relative size of different regions.}
\label{fig:purland}
\end{figure}

\item  Since all states from the accessible region have the same energy probability distribution 
$\{\wkurzGas\}$ (remember (\ref{eq:44})) and the minimum purity state $\Dmin$ is consistent with this 
distribution, all other states within the accessible region that feature $\Pgas\approx \Pmin$ must 
yield reduced local states that are very close to $\Dmin$ (in this context close means in terms of the 
distance measure $\tr{(\dichtematrix^g-\Dmin)^2}$). 
Thus, as long as the trajectory keeps wandering through the compartment filled with those states the 
gas system is locally in a stationary state, equilibrium is reached.

\end{enumerate}
Details can be found in Ref.~\cite{Gemmer2004,GemmerOtte2001}.

%
%
\subsection{Microcanonical Equilibrium}
\label{chap:qt:sec:microcan:5}

The minimal purity for subsystem g under the given constraints is
\begin{equation}
\Pmin = \sum_A \frac{(\wkurzGas)^2}{\EntkurzGas}\;. 
\end{equation}
 
For the \hsav\ of $\Pgas$ in the accessible region we find
\begin{align}
  \label{eq:340}
  \Haver{\Pgas} %
    =& \sum_A\frac{\wahr{2}{A}}{\EntkurzGas}\left(1- %
       \sum_B\wahr{2}{B}\right) \notag\\[3mm]
    &+  \sum_B\frac{\wahr{2}{B}}{\EntkurzCon}\left(1- %
       \sum_A\wahr{2}{A}\right) \notag\\[3mm]
    &+ \sum_{A,B}\frac{\wahr{2}{A}\wahr{2}{B}
                      (\EntkurzGas+\EntkurzCon)}{\EntkurzGas\EntkurzCon+1}\;.
\end{align}
This average is thus a unique function of the invariants $\wkurzGas$, $\wkurzCon$, specified by the initial product state and the degeneracies $\EntkurzGas$, $\EntkurzCon$.

If the degeneracy of the occupied energy levels is large enough so that 
\begin{equation}
  \label{eq:341}
  \EntkurzGas \EntkurzCon + 1 \approx \EntkurzGas \EntkurzCon\,,
\end{equation}
which should hold true for typical thermodynamic systems, (\ref{eq:340}) reduces to
\begin{equation}
  \label{eq:64}
  \Haver{\Pgas} \approx \sum_A \frac{(\wkurzGas)^2}{\EntkurzGas} %
                     +\sum_B \frac{(\wkurzCon)^2}{\EntkurzCon}\;.
\end{equation}
The first sum in this expression is obviously exactly $\Pmin$, so that for systems and initial conditions, in which the second sum is very small, the allowed region almost entirely consists of states for which $\Pgas\approx \Pmin$. 
The second sum will be small if the container system occupies highly degenerate states typical for thermodynamic systems, in which the surrounding is much larger than the considered system. 
This is the set of cases mentioned already in \refsec{chap:qt:sec:microcan:2}: all systems fulfilling this \precondition\ are called now \emph{thermodynamic systems}. 
Thus we can conclude that all states within the accessible region are very close to $\Dmin$ and have approximately the purity $\Pmin$. 
The \densityoperator, which has $\Pgas=\Pmin$ and $\Sgas=\Smax$, and which is consistent with the microcanonical conditions, is unique.
The {\densityoperator}s with $\Pgas\approx \Pmin$ should not deviate much from this one and should therefore 
also have $\Sgas \approx \Smax$, the latter being
\begin{equation}
  \label{eq:65}
  \Smax = -\kBolz \sum_A \wkurzGas
                  \ln\frac{\wkurzGas}{\EntkurzGas}\;.
\end{equation}

%
%
\section{Energy Exchange Conditions}
\label{chap:qt:sec:can}

%
%
\subsection{The Accessible and the Dominant Region}
\label{chap:qt:sec:can:1}

Our approach to the \qmarks{energy exchange conditions} will be based on similar techniques as before. 
The possibility of a partition is still assumed. 
But now there is no further constraint on the interaction $\WW$, since energy is allowed to flow from one \subsystem\ to the other. 
The only constraint for the accessible region therefore derives from the initial state of the full system, and the fact that the probability to find the total system at some energy $\Energy$,
\begin{equation}
  \label{eq:67}
  \WahrscheinlichkeitEnergy := \sum_{A,B/E} \wkurzBei %
       = \sum_{A,B/E} \sum_{a,b} \Abs{\PsiElem{a}{b}{A}{B}}^2 \;,
\end{equation}
should be conserved, where $A,B/E$ stands for: all $A,B$ such that $\EnergyGas+\EnergyCon = E$. 
This constraint is nothing else but the overall \energyconservation.

One could try to repeat the above calculation under the \energyconservation\ constraint, but now it turns out that the average purity over the accessible region is no longer close to the actual minimum purity.
Furthermore, the energy probability distribution of the individual considered system is no longer a constant of motion.
Thus, we proceed in a slightly different way:

\begin{enumerate}

\item Contrary to the microcanonical case the probability to find the gas (container) \subsystem\ at some given energy 
is no longer a constant of motion here. 
But one can prove that there is a predominant distribution, $\{\wkurzDom\}$, which almost all states within the allowed region have in common. 
The subregion formed by these states is called the \qmarks{dominant region}.

\item Having identified the \qmarks{dominant region} we demonstrate that this 
region is by far the biggest subregion in the accessible region of the system 
(see \reffig{fig:purlandcc}).

\begin{figure}
\centering
\psfrag{1}{1}
\psfrag{0}{0}
\psfrag{p_rest}{\tiny $P \approx \Pmin$}
\psfrag{p_min}{\raisebox{-2pt}{\tiny $\Pmin$}}
\psfrag{empty}{}
\psfrag{peq1}{\tiny $P = 1$}
\psfrag{rmin}{$\Dmin$}
\psfrag{AR}{\small Accessible Region}
\psfrag{DR}{\raisebox{-2mm}{\hspace{-5mm}\small Dominant Region}}
\includegraphics[width=5cm]{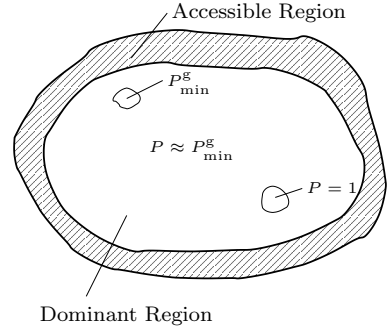}
\caption{Qualitative picture of the purity landscape. In the canonical case the accessible region contains a dominant region which almost entirely fills the accessible region. Within the dominant region, all states feature the same energy probability distribution. Thus all topology from the microcanonical case (\cf ~\reffig{fig:purland}) transfers to the dominant region.}
\label{fig:purlandcc}
\end{figure}

 \item Once the existence of such a dominant region has been established, we can use the results from the microcanonical conditions to argue that almost all states within this dominant region, which is specified by a fixed energy probability distribution for the considered system, feature the maximum local entropy that is consistent with the predominant distribution. 
Based on this analysis we get the equilibrium state of the considered system (see \refsec{chap:qt:sec:can:4}).

\end{enumerate}

Just like in the previous case our subjective lack of knowledge about where to find the system within the accessible 
region is irrelevant.
The reduced local state $\dgas(t)$ as a function of the full state $\ket{\psi(t)}$ should always evolve into a state 
with a fixed probability distribution $\wkurzGas$, and an almost time invariant entropy, which is the maximum entropy 
consistent with this (canonical) distribution. 
Nevertheless, the state of the full system continues to move in \hilbertspace\ with the constant velocity $v$ ($ v \approx $
energy uncertainty).

Again, details can be found in Ref.~\cite{Gemmer2004,GemmerOtte2001}.

%

%
\subsection{The Equilibrium State}
\label{chap:qt:sec:can:4}

Finally, to find the marginal, dominant energy probability distribution $\wkurzDomGas$ of the gas system individually, one has to sum the compound probabilities $\wkurzDom$ over the irrelevant container system to obtain
\begin{align}
  \label{eq:79}
  \wkurzDomGas 
  &= \sum_B \wkurzDom = \sum_{B/E} \frac{\NAB}{\lambda_E} \nonumber\\
  &= \EntGas \sum_{B/E} \frac{\EntCon \WahrscheinlichkeitEnergy} %
                             {\EntEnergy} \;,
\end{align}
where again the sum over $B/E$ denotes a summation under the condition $E=\EnergyGas+\EnergyCon$, and $\EntkurzGas=\EntGas$ and $\EntkurzCon=\EntCon$ are the respective degeneracies.
Since $\EnergyCon=E-\EnergyGas$ is a function of $E$ for fixed $\EnergyGas$ we switch from a summation over $B$ to a summation over $E$,
\begin{equation}
  \label{eq:80}
  \wkurzDomGas = \EntGas \sum_E %
    \frac{\Ent{\text{c}}(E-\EnergyGas)\WahrscheinlichkeitEnergy}{\EntEnergy}\,.
\end{equation}
This is the energy probability distribution for the \gsystem\ that one will find with overwhelming probability for a thermodynamic system. 
Simply by exchanging the indices (up to here everything is symmetric with respect to an exchange of the {\subsystem}s) we find the marginal dominant energy probability distribution for the container system.

So far we have only established the energy probability distributions for almost all states from the accessible 
region, but nothing has been said about entropy, purity, \etc\ 
The equilibrium state is still undetermined. 
Once the trajectory has entered the dominant region, we can assume that the trajectory will practically never leave it, because this region fills almost the whole accessible region of the system.
But since all states within the dominant region feature the same energy probability distribution, motion within the dominant region will never give rise to any further energy exchange between the {\subsystem}s. 
As a consequence the situation is as if the system was controlled by microcanonical conditions.

Therefore, we can take the arguments from \refsec{chap:qt:sec:microcan:5} to identify the equilibrium state. 
Following this idea, we can start with (\ref{eq:64}) and use the dominant energy distribution $\wkurzDom$, finding for the \hsav\ of the purity of the gas
\begin{equation}
  \label{eq:351}
  \Haver{\Pgas} \approx \sum_A \frac{(\wkurzDomGas)^2}{\EntkurzGas} %
                     +  \sum_B \frac{(\wkurzDomCon)^2}{\EntkurzCon}\;.
\end{equation}
Once more it is possible to conclude that the second term (due to the environment) is much smaller than the first one for a sufficiently degenerate environment. 
The first term is exactly the minimum purity of the gas system within the dominant region.
Thus, almost all states from the dominant region will yield approximately the same local gas state too. 
This equilibrium state $\Deq$ is, again, the state of minimum purity 
(maximum entropy) that is consistent with the dominant energy distribution,
\begin{equation}
  \label{eq:85}
  \Deq \approx \sum_{A,a} \frac{\wkurzDomGas}{\EntkurzGas} %
    \Dgas{A}{a}{A}{a} \;.
\end{equation}

One problem remains: the dominant energy probability distribution $\wkurzDomGas$ (\ref{eq:80}) is not independent of the initial state since different energy probability distributions of the local initial state may result in different overall energy probability distributions $\WahrscheinlichkeitEnergy$, and those clearly enter (\ref{eq:80}) and thus even (\ref{eq:85}).
Normally the canonical contact of standard thermodynamics leads to an equilibrium state, which does not depend on the initial state.
The canonical contact turns out to be a special subclass, as we will demonstrate in the next Section.

%
%
\section{Canonical Conditions}
\label{chap:qt:sec:rcan}

For a canonical situation the gas system should relax into the canonical equilibrium 
state, independent of the initial 
conditions.
This behavior can be found, if a further condition is taken into account: a special form of the 
degeneracy of the environment $\EntkurzCon$.

Let us assume an exponential increase of the container degeneracy
\begin{equation}
  \label{eq:82}
  \EntkurzCon = \konstzwei \, \E^{\konsteins \EnergyCon}\;,
\end{equation}
where $\konsteins$, $\konstzwei$ are some constants. Such a degeneracy structure is typical for modular systems \cite{Gemmer2004}.
We start again from (\ref{eq:80}) using (\ref{eq:82}) for the degeneracy of the environment
\begin{equation}
  \label{eq:83}
  \wkurzDomGas = \EntkurzGas\;\E^{-\konsteins \EnergyGas} %
     \sum_{E} \frac{\konstzwei\;\E^{\konsteins E}\WahrscheinlichkeitEnergy} %
                     {\EntEnergy}\;.
\end{equation}
Obviously, the sum does not depend on $A$ at all.
Since $\wkurzDomGas$ has been constructed as some probability distribution it is still normalized by definition.
Therefore the sum has to reduce to a normalizing factor.
Finally we get for the dominant energy probability distribution of the gas system 
\begin{equation}
  \label{eq:354}
  \wkurzDomGas = \frac{\EntkurzGas\;\E^{-\konsteins \EnergyGas}} %
                      {\sum_{A}\EntkurzGas \E^{-\konsteins \EnergyGas}}\;.
\end{equation}
This result is no longer dependent on the initial state!

The energy probability distributions of almost all states from the accessible region consistent with the constraints is then the canonical distribution:
Since the argumentation for the minimal purity state (state of maximal entropy) remains unchanged, the equilibrium state reads now
\begin{equation}
  \label{eq:355}
  \Deq \approx \frac{1}{\sum_{A}\EntkurzGas \E^{-\konsteins \EnergyGas}} %
               \sum_{A,a} \E^{-\konsteins \EnergyGas} \Dgas{A}{a}{A}{a}\,.
\end{equation}
Obviously, this is the well known canonical equilibrium state with the inverse 
temperature $\beta=\alpha$.


%
%
\section{Temperature}
\label{chap:temperature}

One could claim that temperature should only be defined for equilibrium and thus there was no need to define it as a function of the \microstate. 
Based on this reasoning temperature would then simply be defined as 
\begin{equation}
  \label{eq:228}
    \frac{1}{\kBolz\Temp} = \pop{\Entropy}{\Energy} 
              = \pop{}{\Energy} \ln \ZDichte{\Energy}
              = \frac{1}{\ZDichte{\Energy}} \,
                \pop{\ZDichte{\Energy}}{\Energy}\;,
\end{equation}
with $\ZDichte{\Energy}$ being the state density.
In this way one would neglect all dynamical aspects (see \cite{Weidlich1976}), since this definition is based on the \hamiltonian\ of the system rather than on its state.
Strictly speaking, this definition would exclude all situations in which temperature appeares as a function of time or space, because those are \nonequilibrium\ situations. 

A quantity like temperature is essentially determined by two properties: It should take on the same value for two systems in energy exchanging contact, and if the energy of a system is changed without changing its volume, it should be a measure for the energy change per entropy change.

Most definitions rely on the second property. 
\name{Maxwell} connected the mean kinetic energy of a classical particle with temperature. 
In the canonical ensemble (\name{Boltzmann} distribution) it is guaranteed that the energy change per entropy change equals temperature. 
And the ensemble mean of the kinetic energy of a particle equals $\kBolz\Temp$ in this case. 
Thus, if ergodicity is assumed, \ie, if the time average equals the ensemble average, temperature may indeed be defined as the time averaged kinetic energy. 
Similar approaches have been proposed on the basis of the microcanonical ensemble 
\cite{Rickayzen2001,Rugh1997}.
However, if temperature was given by a time average over an observable, the proper averaging time remains open 
and thus the question on what minimum timescale temperature may be defined.
Furthermore, this definition was entirely based on ergodicity. 
Nevertheless, it allows, at least to some extent, for an investigation of processes, in which temperature varies in time and/or space, since that definition is not necessarily restricted to full equilibrium.

To avoid those problems of standard temperature definitions, we present here yet another, entirely 
quantum mechanical definition.

%
%
\section{Definition of Spectral Temperature}
\label{chap:temperature:sec:spec}

We define the inverse spectral temperature as \cite{Gemmer2004}
\begin{align}
  \label{eq:150}
  &\frac{1}{\kBolz \Temp}:= %
     -\left(1-\frac{\wahrwn{0}+\wahrwn{M}}{2}\right)^{-1} \notag\\
     &\sum_{i=1}^M \left(\frac{\wahrwn{i}+\wahrwn{i-1}}{2}\right) %
      \frac{\ln(\wahrwn{i}/\wahrwn{i-1})-\ln(\Entn{i}/\Entn{i-1})} %
           {\Energyn{i}-\Energyn{i-1}}\;,
\end{align}
where $\wahrwn{i}$ is the probability to find the quantum system at the energy $\Energyn{i}$, $M$ is the number of the highest energy level $\Energyn{M}$, while the lowest one is labeled $\Energyn{0}$. 
This formula is motivated by the following idea: For a two level system it seems plausible to define temperature just from the energy probability distribution and the degrees of degeneracy as 
\begin{equation}
  \label{eq:167}
  \frac{\wahrwn{1} \Entn{0}}{\wahrwn{0} \Entn{1}} = %
  \expfkt{-\frac{\Energyn{1}-\Energyn{0}}{\kBolz\Temp}}\;.
\end{equation}
The definition (\ref{eq:150}) results if one groups the energy levels of a \multilevel\ system into neighboring pairs, to each of which a \qmarks{temperature} is assigned via the above formula, weighted by the average probability for each pair to be occupied. 
This definition obviously depends only on the energy probability distribution and the spectrum of a system. 
It thus cannot change in time for an isolated system, and it is always defined, independent of whether or not the system is in an equilibrium state. 
Thus there should be many systems or situations with such a temperature, which do not exhibit thermodynamic properties at all. 
The latter will only show up in equilibrium situations or close to those.

If the spectrum of a system was very dense and if it was possible to describe the energy probability distribution, $\{\wahrwn{i}\}$, as well as the degrees of degeneracy, $\{\Entn{i}\}$, by smooth continuous functions ($\WahrscheinlichkeitEnergy,\EntEnergy$) with a well defined derivative, (\ref{eq:150}) could be approximated by
\begin{align}
  \label{eq:168}
  &\frac{1}{\kBolz\Temp} \approx \notag\\%
      &-\int_0^{\Emax} \WahrscheinlichkeitEnergy %
           \left(\dod{}{\Energy} \ln\WahrscheinlichkeitEnergy%
                -\dod{}{\Energy}\right) \ln\EntEnergy%
       \D\Energy\;.
\end{align}
This can further be simplified by integrating the first term to yield
\begin{align}
  \label{eq:169}
  \frac{1}{\kBolz\Temp} %
     \approx& \Wahrscheinlichkeit{0} - \Wahrscheinlichkeit{\Emax}\notag\\
       &+ \int_0^{\Emax} \frac{\WahrscheinlichkeitEnergy}{\EntEnergy} %
       \;\dod{\EntEnergy}{\Energy}\;\D\Energy\;.
\end{align}
Since for larger systems typically neither the lowest nor the highest energy level is occupied with considerable probability (if the spectra are finite at all), it is the last term on the right hand side of (\ref{eq:169}), that basically matters. 
This term can be interpreted as the average over the standard, system based, rather than \microstate\ based definition of the inverse temperature.


%
%
\section{Equilibrium Properties of Model Systems}
\label{chap:num}

We now turn to some numerical data, based on a certain type of models. 
These models are still rather abstract and may thus be viewed as models for a whole class of 
bipartite systems.
The subsystems g, c are specified by their respective spectra only, or, rather, those respective parts which play any role at all under the condition of \energyconservation.
What sort of physical structure could give rise to those spectra is not considered here, since it turns out to be irrelevant to some extent. 

The spectra are translated into discrete diagonal matrix {\hamiltonian}s that describe the decoupled \bipartite\ system. 
These can be chosen to be diagonal, without any loss of generality, since any such system may be analyzed in the energy eigenbasis of its decoupled parts. 
The form of the interaction depends on the concrete physical subsystems and their interactions. 
But since the \qmarks{guess} is that for the quantities considered here (entropy, occupation probabilities, \etc) the concrete form of the interaction should not matter, the interaction is taken as some random matrix thus avoiding any bias. 
The interaction has to be \qmarks{small}.
In this way it is hoped that we get models that are as \qmarks{typical} for general thermodynamic 
situations as possible.
Many of the situations analyzed in this Chapter are very similar to those treated within the context of 
quantum {\masterequation}s.
But note that in order to apply the theories at hand to those systems, neither a Markovian nor a Born assumption has to hold. 
We simply solve the respective Schr\"odinger equation in finite dimensional space.
 
%
%
\subsection{Entropy under Microcanonical Conditions}
\label{chap:num:sec:mc}

\begin{figure}
\centering
\psfrag{E}{\hspace{2mm}$\Delta E$}
\psfrag{S}[c]{$\bigotimes$}
\psfrag{WW}[c]{$\WW$}
\psfrag{N}{$\Ent{c}=50$}
\psfrag{system}{\hspace{-1mm}system}
\psfrag{environment}{\hspace{-2mm}environment}
\includegraphics[width=5cm]{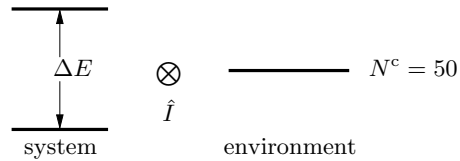}
\caption{Microcanonical scenario: A \nondegenerate\ two-level-system (gas) is weakly coupled to a system with one energy level of degeneracy $\Ent{c}=50$. This is a model for a system in contact with a much larger environment such that no energy can be exchanged between the system and environment.}
\label{fig:micspec}
\end{figure}

All data in our first example refer to a situation depicted in \reffig{fig:micspec} (\cf\ \cite{BorowskiGemmer2003}). 
The \qmarks{gas} (the system under consideration) consists of a \twolevelsystem, both levels being \nondegenerate\ ($\Entartung{g}{0} = \Entartung{g}{1} = 1$), while the \qmarks{container} (the environment) consists of just one energy level with degeneracy $\Ent{c}=50$. 
This is necessarily a microcanonical situation regardless of the interaction $\WW$. 
The container cannot absorb any energy, therefore energy cannot be exchanged between the systems. 
In this situation the probabilities to find the gas system in the ground (excited) state $\wahr{\text{g}}{0}$ ($\wahr{\text{g}}{1}$) are conserved quantities and in this example chosen as 
\begin{equation}
  \label{eq:115}
  \wahr{\text{g}}{0} = 0.15\;,\quad \wahr{\text{g}}{1} = 0.85\;.
\end{equation}
As described in \refsec{chap:qt:sec:microcan}, the \hsav\ of the purity of the gas system is under condiction (\ref{eq:341}) given according to (\ref{eq:64}) by $\Haver{\Pgas}=0.765$.
The corresponding minimum purity (\ref{eq:157}) is $\Pmin= 0.745$.
As explained in \refsec{chap:qt:sec:microcan}, we find here
\begin{equation}
  \label{eq:119}
  \Haver{\Pgas} \approx \Pmin\;,
\end{equation}
a situation, in which almost the entire accessible region would be filled with the compartment containing only states of almost maximum local entropy.

To confirm this expectation, a set of random states, uniformly distributed over the accessible region, has been generated. 
Their local entropies have been calculated and sorted into a histogram. 
Since those states are distributed uniformly over the accessible region, the number of states in any \qmarks{entropy bin} reflects the relative size of the respective \hilbertspace\ compartment.
 
\begin{figure}
\centering
\psfrag{n}{\hspace*{-2cm}\tiny Size of \hilbertspace\ compartment}
\psfrag{S}{\raisebox{-3pt}{$S[\kBolz]$}}
\includegraphics[height=7cm,angle=-90]{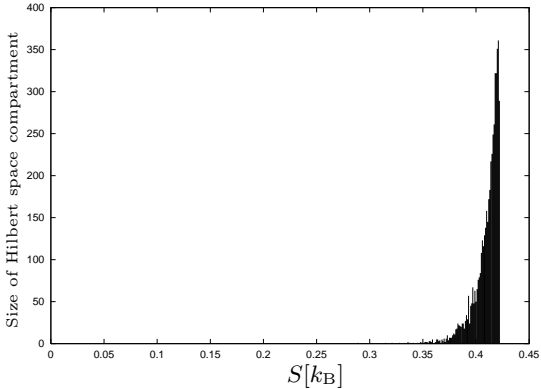}
\caption{Relative size of \hilbertspace\ compartments: this histogram shows the relative frequency of states with a given local entropy $\Entropy$, among all states from the accessible region. In this case the maximum possible entropy is $\Smax=0.423\,\kBolz$. Obviously, almost all states feature entropies close to the maximum.}
\label{fig:micsta}
\end{figure}

The histogram is shown in  \reffig{fig:micsta}. 
The maximum local entropy in this case is $\Smax = 0.423\,\kBolz$. 
Obviously, almost all states have local entropies close to $\Smax$. 
Thus compartments corresponding to entropies of, say, $\Sgas > 0.4\,\kBolz$ indeed fill almost the entire accessible region, just as theory predicts. 
Local pure states ($\Sgas=0$) are practically of measure zero.

In order to examine the dynamics, a coupling $\WW$ is introduced. 
To keep the concrete example as general as possible, $\WW$ has been chosen as a random matrix in the basis of the energy eigenstates of the uncoupled system, with \Gaussian\ distributed real and imaginary parts of the matrix elements of zero mean and a standard deviation of
\begin{equation}
  \label{eq:120}
  \Delta I=0.01 \Delta E\;.
\end{equation}
This coupling is weak, compared to the \hamiltonian\ of the uncoupled system.
Therefore the respective interaction cannot contain much energy.
The spectrum of the system (see \reffig{fig:micspec}) does not change significantly due to the coupling, and afterall the environment is not able to absorbe energy. 

\begin{figure}
\centering
\psfrag{Zeit}{\raisebox{-3pt}{$t[\frac{\hbar}{\Delta E}]$}}
\psfrag{S}{$S[\kBolz]$}
\includegraphics[height=7cm,angle=-90]{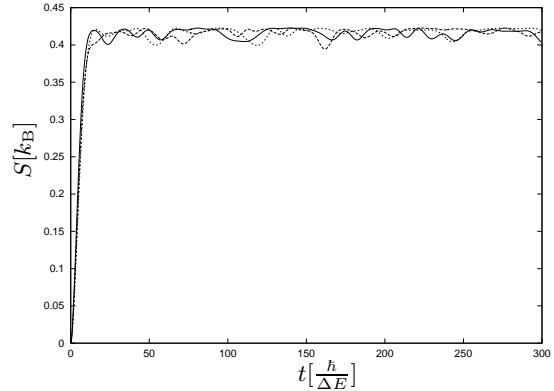}
\caption{Evolution of the local entropy for different initial states. A universal state of maximum entropy (equilibrium) is reached, independent of the initial state.}
\label{fig:micev}
\end{figure}

Now the \sgleichung\ for this system, including a realization of the interaction, has been solved for initial states consistent with (\ref{eq:115}). 
Then the local entropy at each time has been calculated, thus getting a picture of the entropy evolution. 
The result is shown in \reffig{fig:micev}. 
Obviously the entropy approaches $\Smax$ within a reasonable time, regardless of the concrete initial state. 
Thus the tendency towards equilibrium is obvious. 
The concrete form of the interaction $\WW$ only influences the details of this evolution, the equilibrium value is always the same. 
If the interaction is chosen to be weaker, the time scale on which equilibrium is reached gets longer, but, eventually the same maximum entropy will be reached in any case.

%
%
\subsection{Occupation Probabilities under Canonical Conditions}
\label{chap:num:sec:cc}

\begin{figure}
\centering
\psfrag{E}{\hspace{1mm}$\Delta E$}
\psfrag{S}[c]{$\bigotimes$}
\psfrag{WW}[c]{$\WW$}
\psfrag{system}{\hspace{-1mm}system}
\psfrag{environment}{\hspace{-1mm}environment}
\psfrag{N1}{$\Entartung{c}{0}=50$}
\psfrag{N2}{$\Entartung{c}{1}=100$}
\psfrag{N3}{$\Entartung{c}{2}=200$}
\includegraphics[width=5cm]{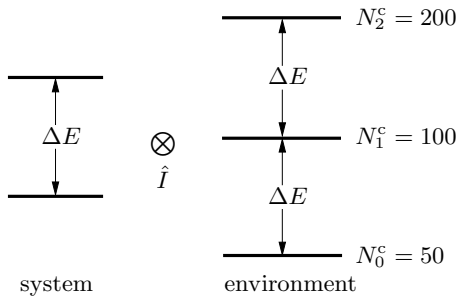}
\caption{Canonical scenario: A \twolevel\ gas system is weakly coupled to a three level environment, such that energy can be exchanged. The exponential degeneracy scheme of the container system guarantees a full independence of the equilibrium state from the initial state.}
\label{fig:kanspec}
\end{figure}

The second model analyzed numerically is depicted in \reffig{fig:kanspec}. 
The considered (gas) system, again, consists only of a \nondegenerate\ \twolevelsystem. 
The environment (container) in this case is a \threelevelsystem\ with an exponential \qmarks{state density}: $\Entartung{c}{B} = 50 \cdot 2^B$ with $B=0,1,2$. 
This has been chosen since theory predicts for such a degeneracy scheme of the environment an equilibrium state of the gas system, which should be independent of its initial state (see (\ref{eq:83})).
If we restrict ourselves to initial states featuring arbitrary states for the gas system but container states that only occupy the intermediate energy level, no other container levels except for those given could be reached, even if they were present: This is due to \energyconservation\ and holds for the limit of weak interactions $\WW$.

In this case the model can also be seen to represent a situation with a much larger environment and we find from (\ref{eq:354})
\begin{equation}
  \label{eq:121}
  \WahrscheinlichkeitDomGas[0] = \frac{2}{3}\;, \quad%
  \WahrscheinlichkeitDomGas[1] = \frac{1}{3}\;.
\end{equation}
To keep the situation as general as possible, $\WW$ was, like in \refsec{chap:num:sec:mc}, chosen to be a matrix with random \Gaussian\ distributed entries in the basis of the eigenstates of the uncoupled system, but now with energy transfer allowed between the subsystems.

\begin{figure}
\centering
\psfrag{Zeit}{\raisebox{-3pt}{$t[\frac{\hbar}{\Delta E}]$}}
\psfrag{rho^g_00}{$\WahrscheinlichkeitGas[0]$}
\includegraphics[height=7cm,angle=-90]{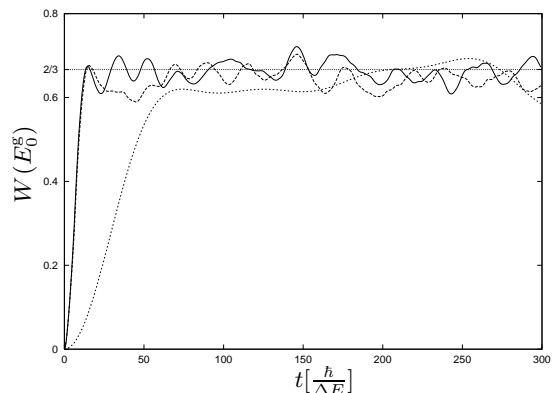}
\caption{Evolution of the ground level occupation probability for three different random interactions. The dotted line corresponds to a weaker interaction. Even in this case the same equilibrium value, $\wkurzDomGas=\frac{2}{3}$, is approached, only on a longer timescale.}
\label{fig:kanproev}
\end{figure}

For this system the \sgleichung\ has been solved and the evolution of the probability to find the gas system in its ground state, $\WahrscheinlichkeitGas[0]$ is displayed in \reffig{fig:kanproev}. 
The different curves correspond to different interaction strengths, given by the standard deviation of the distribution of the matrix elements of $\WW$, $\Delta I$:
\begin{equation}
  \label{eq:122}
  \Delta I_{\text{solid, dashed}}=0.0075\Delta E \;,\quad %
  \Delta I_{\text{dotted}}=0.002\Delta E\;.
\end{equation}
Obviously, the equilibrium value of $\WahrscheinlichkeitDomGas[0]=2/3$ is reached independently of the concrete interaction $\WW$. 
Within the weak coupling limit the interaction strength only influences the timescale on which equilibrium is reached.

\begin{figure}
\centering
\psfrag{Zeit}{\raisebox{-3pt}{$t[\frac{\hbar}{\Delta E}]$}}
\psfrag{rho^g_00}{$\WahrscheinlichkeitGas[0]$}
\includegraphics[height=7cm,angle=-90]{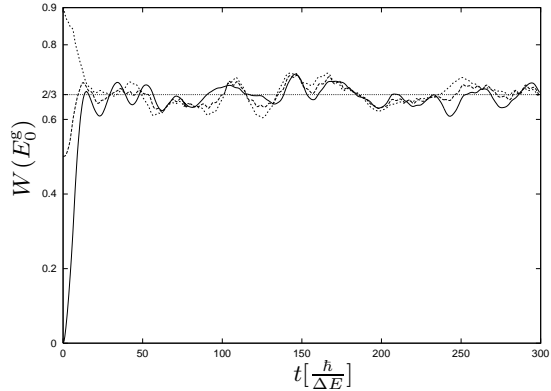}
\caption{Evolution of the ground level occupation probability for different initial states. The theoretically predicted equilibrium value is reached, independent of the initial states, as expected for canonical conditions. }
\label{fig:kanproevs}
\end{figure}

Figure \ref{fig:kanproevs} displays the evolution of the same probability, $\WahrscheinlichkeitGas[0]$, but now for different initial states, featuring different probabilities for the groundstate, as can be seen in the figure at $t=0$. 
The equilibrium value is reached for any such evolution, regardless of the special initial state, thus we confirm the effective attractor behavior typical for thermodynamics.

\begin{figure}
\centering
\psfrag{Zeit}{\raisebox{-3pt}{$t[\frac{\hbar}{\Delta E}]$}}
\psfrag{S^g}{$S[\kBolz]$}
\includegraphics[width=7cm]{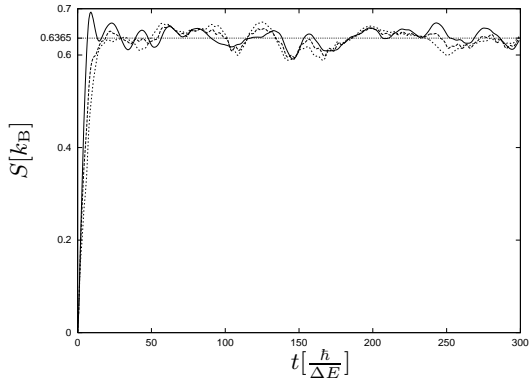}
\caption{Evolution of the local entropy for different initial states. $S=0.637\,\kBolz$ is the maximum entropy that is consistent with the equilibrium energy probabilities. This maximum entropy state is reached in all cases.}
\label{fig:kanentrosev}
\end{figure}

Figure \ref{fig:kanentrosev} shows the evolution of the local entropy of the gas system for the same three initial states as used for \reffig{fig:kanproevs}.
\begin{figure}
\centering
\psfrag{E}{\small $\Delta E$}
\psfrag{S}{$\bigotimes$}
\psfrag{WW}{\hspace{2mm}\raisebox{-3pt}{$\WW$}}
\psfrag{system}{\hspace{-3mm}system}
\psfrag{environment}{\hspace{-2mm}environment}
\psfrag{N1}{$\Entartung{c}{0}=6$}
\psfrag{N2}{$\Entartung{c}{1}=12$}
\psfrag{N3}{$\Entartung{c}{2}=24$}
\psfrag{N4}{$\Entartung{c}{3}=48$}
\psfrag{N5}{$\Entartung{c}{4}=96$}
\includegraphics[width=3cm]{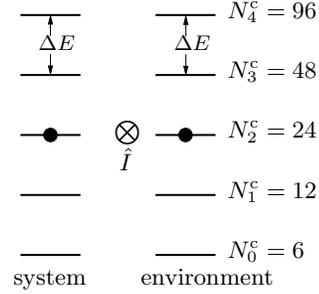}
\caption{Canonical \multilevel\ scenario: A \fivelevel\ gas system is weakly coupled to a five level container system with an exponential degeneracy scheme, such that energy may be exchanged. Black dots symbolize the initial state. This \setup\ should lead to a \name{Boltzmann} distribution.}
\label{fig:kanspec5}
\end{figure}

The maximum entropy, consistent with the equilibrium value of the energy probabilities, is $\Smax=0.637\,\kBolz$. 
This is also the value one finds, if one maximizes entropy for fixed mean energy (\name{Jaynes'} principle \cite{Jaynes1957II}). 
Obviously, this value is reached for any initial state during the concrete dynamics of this model. 
This supports the validity of (\ref{eq:85}), which states that the \densitymatrix\ of the equilibrium state is diagonal in the basis of the local energy eigenstates.

\begin{figure}
\centering
\psfrag{Zeit}{\raisebox{-3pt}{$t[\frac{\hbar}{\Delta E}]$}}
\psfrag{rho^g_ii}{$\WahrscheinlichkeitGas$}
\includegraphics[height=7cm,angle=-90]{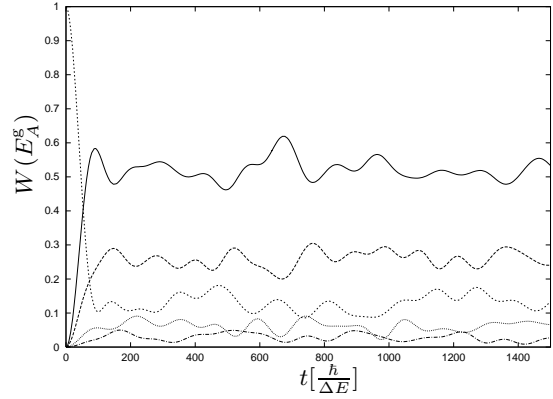}
\caption{Evolution of the energy occupation probabilities. After some relaxation time a Boltzmann distribution is reached. Each probability is twice as high as the one for the next higher energy level, as theory predicts.}
\label{fig:kanprosev}
\end{figure}

To analyze the formation of a full Boltzmann distribution, we turn to the system depicted in \reffig{fig:kanspec5}. 
Here the \qmarks{gas} system is a \nondegenerate\, equidistant \fivelevelsystem\, the container system a \fivelevelsystem\ with degeneracies $\Entartung{c}{B}= 6 \cdot 2^B$ ($B=0,\dots,4$), which should lead to a Boltzmann distribution. 
We restrict ourselves to initial states, where for both subsystems only the intermediate energy level is occupied (symbolized by the black dots in \reffig{fig:kanspec5}). 
Due to \energyconservation\ other states of the container system would not play any role in this case even if they were present, just like in the previous model. 
Figure \ref{fig:kanprosev} shows the probabilities of the different energy levels to be occupied $\WahrscheinlichkeitGas$. 
While the gas system starts in the intermediate (third) energy level, soon a \name{Boltzmann} distribution develops: 
Obviously, each probability becomes twice as high as the one for the level above. 
This is exactly what theory predicts (see (\ref{eq:80})) for the environment degeneracy scheme under consideration. 

%
%
\subsection{Probability Fluctuations}
\label{chap:num:sec:fluc}

Finally, we consider fluctuations in time due to finite size effects imposed by the environment.
For this purpose a system almost like the one depicted 
in \reffig{fig:kanspec} is analyzed, but now with a degeneracy
scheme given by
\begin{equation}
  \label{eq:123}
  \Entartung{c}{B} = \frac{\Entartung{c}{1}}{2} \cdot 2^B \;.
\end{equation}
The ratios between the degrees of degeneracy of the different container levels are thus the same as for the system sketched in \reffig{fig:kanspec}, but the overall size of the container system is tunable by $\Entartung{c}{1}$.
For various $\Entartung{c}{1}$, the \sgleichung\ has been solved numerically, and the following measure of the fluctuations of the occupation probability of the ground level of the gas system has been computed
\begin{align}
  \label{eq:124}
  \Delta_t^2 \wahr{\text{g}}{0} := %
  \frac{1}{t_f-t_i} %
   &\left( \int_{t_i}^{t_f}\big(\wahr{\text{g}}{0}(t)\big)^2 \,\D t \right.
   \notag\\
  -&\left.\left( \int_{t_i}^{t_f} \wahr{\text{g}}{0}(t)\,\D t \right)^2\right)
  \;,
\end{align}
for initial states with
\begin{equation}
  \label{eq:125}
  \wahr{\text{g}}{0}(0) = 0.2\;,\quad 
  \wahr{\text{g}}{1}(0) = 0.8\;.
\end{equation}

\begin{figure}
\centering
\psfrag{(Delta rho)^2}{$\Delta_t^2 \wahr{\text{g}}{0}$}
\psfrag{n^c(E^c_2)}{\raisebox{-3pt}{$\Entartung{c}{1}$}}
\psfrag{Simulationen}{\hspace{-6mm}\tiny simulation}
\psfrag{least square fit}{\hspace{0mm}\tiny fit}
\includegraphics[height=7cm,angle=-90]{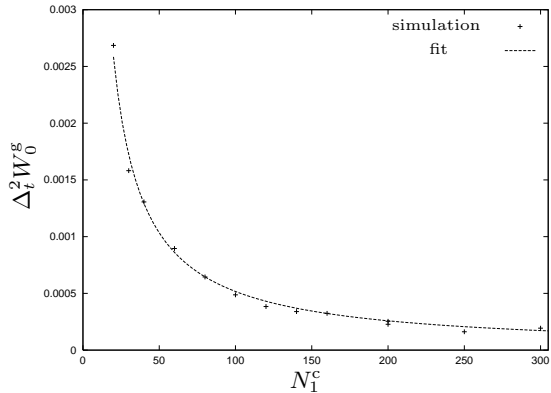}
\caption{Fluctuations of the probability for the considered system to be in the ground state $\Delta_t \wahr{\text{g}}{0}$, in dependence on the number of eigenstates of the environment system $\Entartung{c}{1}$. Obviously, the fluctuations go down, with increasing environment.}
\label{fig:varianzen}
\end{figure}

Figure \ref{fig:varianzen} shows the dependence of the size of these fluctuations on the container system size $\Entartung{c}{1}$. 
The small crosses are the computed data points, the dashed line is a least square fit to a function 
proportional to $1/ \Entartung{c}{1}$,
\begin{equation}
  \label{eq:126}
  \Delta_t \wahr{\text{g}}{0} =
  \sqrt{\frac{0.053}{\Entartung{c}{1}}}\;.
\end{equation}
This fit is in very good agreement with the theoretical expectation, although the trajectories are not ergodic.

%
\section{Summary and Conclusions}

The reduction of thermodynamics to an underlying theory has challenged physicists for more than a century. So far such
attempts have met with partial success only. In this context an interesting and, as we believe, 
consistent route is offered by quantum thermodynamics.

Quantum thermodynamics as a concept has not yet a generally accepted meaning (see, e.g.\ \cite{Hoffmann2001} for a different opinion).
Our version has been introduced as a special field within decoherence theory. It makes use of a special
methodology, which is based on statistical techniques to investigate the structure of a product Hilbert space under
certain constraints. These constraints, though, are by themselves emergent properties of the system environment model.

Within this quantum thermodynamics we are able to show that thermodynamics is already contained in the deterministic Schr\"odinger
dynamics. Fingerprints of equilibrium behavior show up already in suprisingly small bipartite quantum systems. As a
consequence thermodynamics need not be considered as a final remedy after all detailed microscopic 
descriptions have failed: Deterministic Schr\"odinger dynamics of the whole and (apparent) relaxation behavior of the
embedded part is no contradiction at all.

Such a picture should also give new insight on thermodynamic applications in the
nano-regime \cite{Ritort2004}. This expectation is corroborated by the fact that pertinent models then have to be on the 
quantum level anyway, 
while the conditions for a thermodynamic description can be made explicit, as this is part of the complete quantum
description of system and environment \cite{Nieuwenhuizen2002}.

{\bf Ackknowledgement:} We thank the Deutsche Forschungsgemeinschaft for financial support.

\end{document}